\documentclass[12pt]{article}
\usepackage{graphicx}

\newcommand\pubnumber{CMS CR-2017/369}
\newcommand\pubdate{\today}

\def\institute{Institute of Experimental Particle Physics\\
Karlsruhe Institute of Technology, 76131 Karlsruhe, GERMANY}

\def\Title#1{\begin{center} {\Large #1 } \end{center}}
\def\Author#1{\begin{center}{ \sc #1} \end{center}}
\def\Address#1{\begin{center}{ \it #1} \end{center}}

\newcommand\pubblock{\rightline{\begin{tabular}{l} \pubnumber\\
         \pubdate  \end{tabular}}}
\newenvironment{Abstract}{\begin{quotation}  }{\end{quotation}}
\newenvironment{Presented}{\begin{quotation} \begin{center} 
             PRESENTED AT\end{center}\bigskip 
      \begin{center}\begin{large}}{\end{large}\end{center} \end{quotation}}

\def\beq{\begin{equation}}
\def\eeq#1{\label{#1}\end{equation}}
\def\eeqn{\end{equation}}

\def\beqa{\begin{eqnarray}}
\def\eeqa#1{\label{#1}\end{eqnarray}}
\def\eeqan{\end{eqnarray}}

\let\bar=\overbar

\def\Dslash{\not{\hbox{\kern-4pt $D$}}}
\def\dslash{\not{\hbox{\kern-2pt $\del$}}}

\def\msb{{\bar{\ssstyle M \kern -1pt S}}}

\begin{document}
\begin{titlepage}
\pubblock

\vfill
\Title{Search for $\mathrm{H}\to\mathrm{b\bar{b}}$ in association with a single top quark\\at CMS}
\vfill
\Author{Denise M\"uller\\on behalf of the CMS Collaboration}
\Address{\institute}
\vfill
\begin{Abstract}
A search for the production of a Higgs boson in association with a single top quark (tH) is performed using the decay $\mathrm{H}\to\mathrm{b\bar{b}}$. This Higgs boson production mode is sensitive to the coupling modifiers of the Higgs boson to top quarks ($\kappa_\mathrm{t}$) and to weak gauge bosons ($\kappa_\mathrm{V}$) and thus can be used to lift the degeneracy regarding the sign of the top quark Yukawa coupling. The 2015 pp collisions data at a center-of-mass energy of 13~TeV are analyzed. Boosted decision trees are used to reconstruct and classify the events. Upper limits on the tH production cross section are determined at 51 different points in the $\kappa_\mathrm{t}-\kappa_\mathrm{V}$ plane.
\end{Abstract}
\vfill
\begin{Presented}
$10^\mathrm{th}$ International Workshop on Top Quark Physics\\
Braga, Portugal,  September 17--22, 2017
\end{Presented}
\vfill
\end{titlepage}
\def\thefootnote{\fnsymbol{footnote}}
\setcounter{footnote}{0}

\section{Introduction}

\begin{figure}[htb]
\centering
\includegraphics[width=0.4\textwidth]{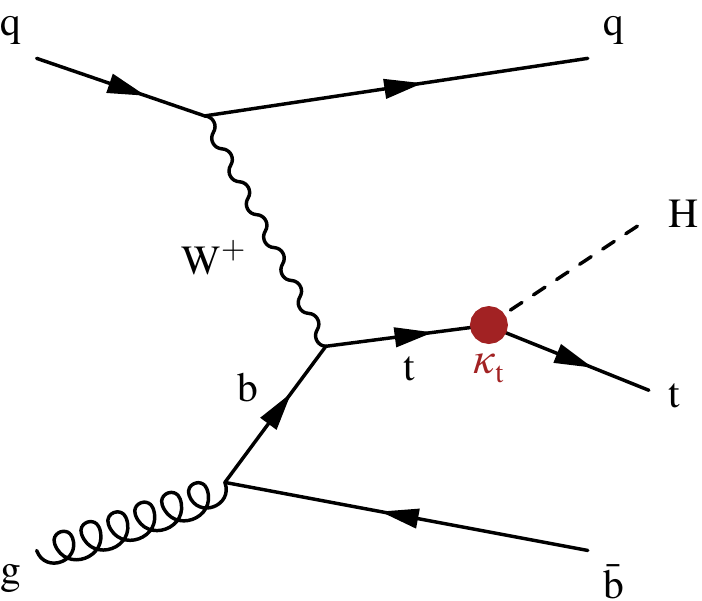}
\hspace*{1.2em}
\includegraphics[width=0.4\textwidth]{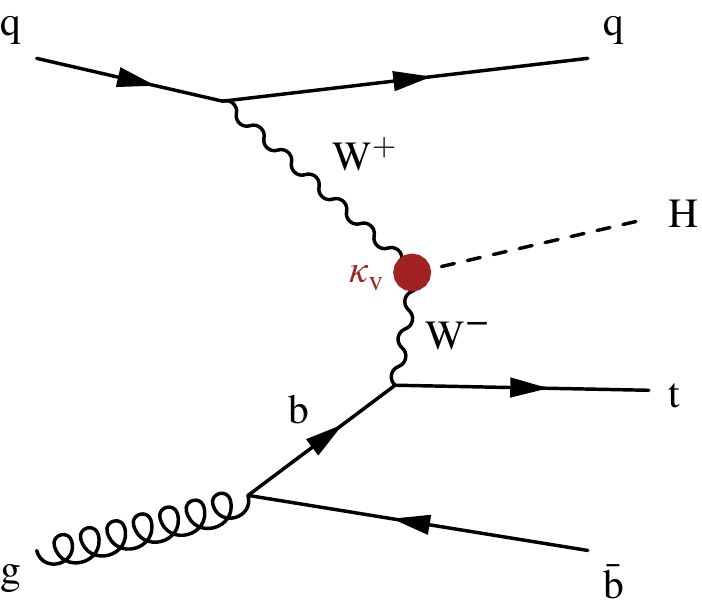}
\caption{Feynman diagrams for the associated production of a single top quark and a Higgs boson in the $t$ channel. Left figure: Higgs boson emitted from the top quark. Right figure: Higgs boson emitted from the intermediate W boson.}
\label{fig:feynman}
\end{figure}

In contrast to most other measurements, the rate of the production of a single top quark in association with a Higgs boson (tH production) is sensitive not only to the magnitude but also to the sign of the Yukawa coupling of the top quark. In this production mode, the single top quark is produced via the $t$ channel or via the associated production with a W boson. Due to its small cross section, the $s$-channel production is negligible. The Higgs boson can be emitted either from the top quark or the intermediate W boson (see Fig.~\ref{fig:feynman}) and the amplitudes of these two possibilities interfere. The resulting amplitude depends on the ratios of actual coupling strengths to the standard model (SM) predictions for the Higgs-top coupling ($\kappa_\mathrm{t}$) and for the coupling of the Higgs boson to vector bosons ($\kappa_\mathrm{V}$), given by $\mathcal{A} \propto (\kappa_\mathrm{t}-\kappa_\mathrm{V})$. A scan over 51 different values of $\kappa_\mathrm{t}$ and $\kappa_\mathrm{V}$, suitable for a test of Higgs boson couplings, is provided. Furthermore, physics beyond the SM can be potentially discovered by the evidence of anomalous couplings. The scenario with a negative Yukawa coupling of the top quark ($\kappa_\mathrm{t} = -1.0$, $\kappa_\mathrm{V} = +1.0$) -- called inverted top coupling (ITC) scenario -- differs only in the sign of $\kappa_\mathrm{t}$ from the SM case ($\kappa_\mathrm{t} = +1.0$, $\kappa_\mathrm{V} = +1.0$). The ITC case is particularly relevant for the analysis, as it has a significantly higher cross section. The increase of the cross section is caused by the constructive interference of the amplitudes of the two emission possibilities of the Higgs boson. In the SM case, a destructive interference of the amplitudes occurs. In this analysis, 51 different points in the $\kappa_\mathrm{t} - \kappa_\mathrm{V}$ plane with $-3.0 \leq \kappa_\mathrm{t} \leq +3.0$ and $\kappa_\mathrm{V} = \{+0.5, +1.0, +1.5\}$ are investigated.

\section{Event Reconstruction and Classification}

The analyzed data set consists of events recorded with the Compact Muon Solenoid (CMS) experiment~\cite{det} during Run II of the Large Hadron Collider (LHC) in 2015, corresponding to an integrated luminosity of $2.3\,\mathrm{fb}^{-1}$. Events are selected which contain exactly one isolated lepton (muon or electron), three or four b-tagged jets and at least one untagged jet. All jets are required to have $p_\mathrm{T}>30\,\mathrm{GeV}$ ($|\eta|<2.4$) and $p_\mathrm{T}>40\,\mathrm{GeV}$ ($|\eta|\geq 2.4$) respectively. Additionally, a cut on the missing transverse energy is applied: ${\not\mathrel{E}}_\mathrm{T}>45\,\mathrm{GeV}$ (electron channel) and ${\not\mathrel{E}}_\mathrm{T}>35\,\mathrm{GeV}$ (muon channel). According to the number of b-tagged jets, two independent signal regions are defined, namely the 3 tag and 4 tag regions.

Each event is reconstructed under the signal hypothesis and under the $\mathrm{t\bar{t}}$ hypothesis, as the $\mathrm{t\bar{t}}$ process is the dominant background of this analysis. In both reconstructions, the jets of an event need to be assigned to the four final-state quarks. Hence, boosted decision trees (BDTs) are trained to separate between correct and random jet-to-quark assignments. In the application of the BDTs, all possible jet permutations are considered. For each permutation, the BDT score is calculated, and the assignment which yields the highest BDT score is chosen. As the kinematics of the tH process depend on $\kappa_\mathrm{t}$ and $\kappa_\mathrm{V}$, 51 different BDTs need to be employed for the tH reconstruction. In case of the $\mathrm{t\bar{t}}$ reconstruction, only one BDT is trained.

\begin{figure}[h!]
\centering
\includegraphics[width=0.40\textwidth]{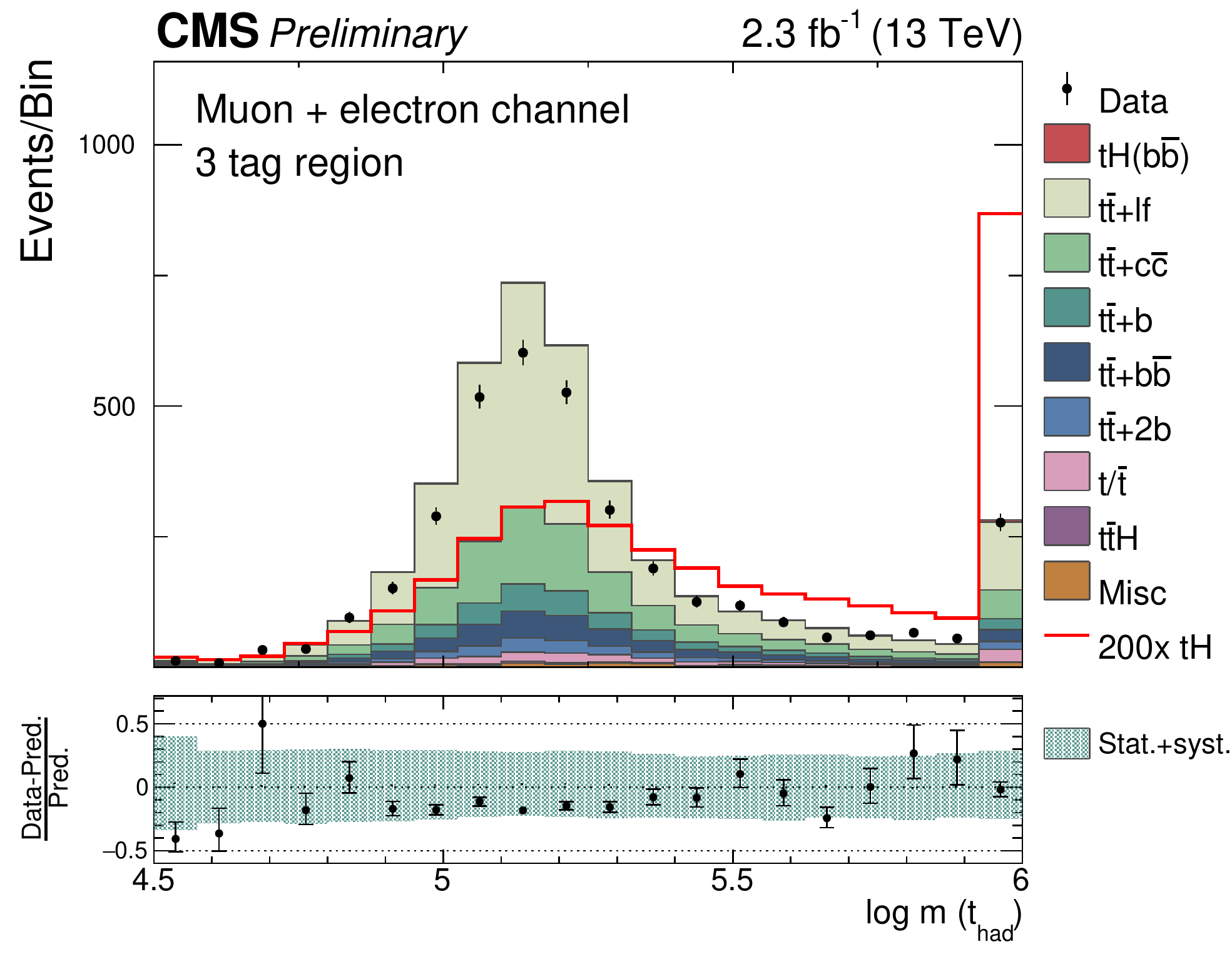}
\hspace*{1.2em}
\includegraphics[width=0.40\textwidth]{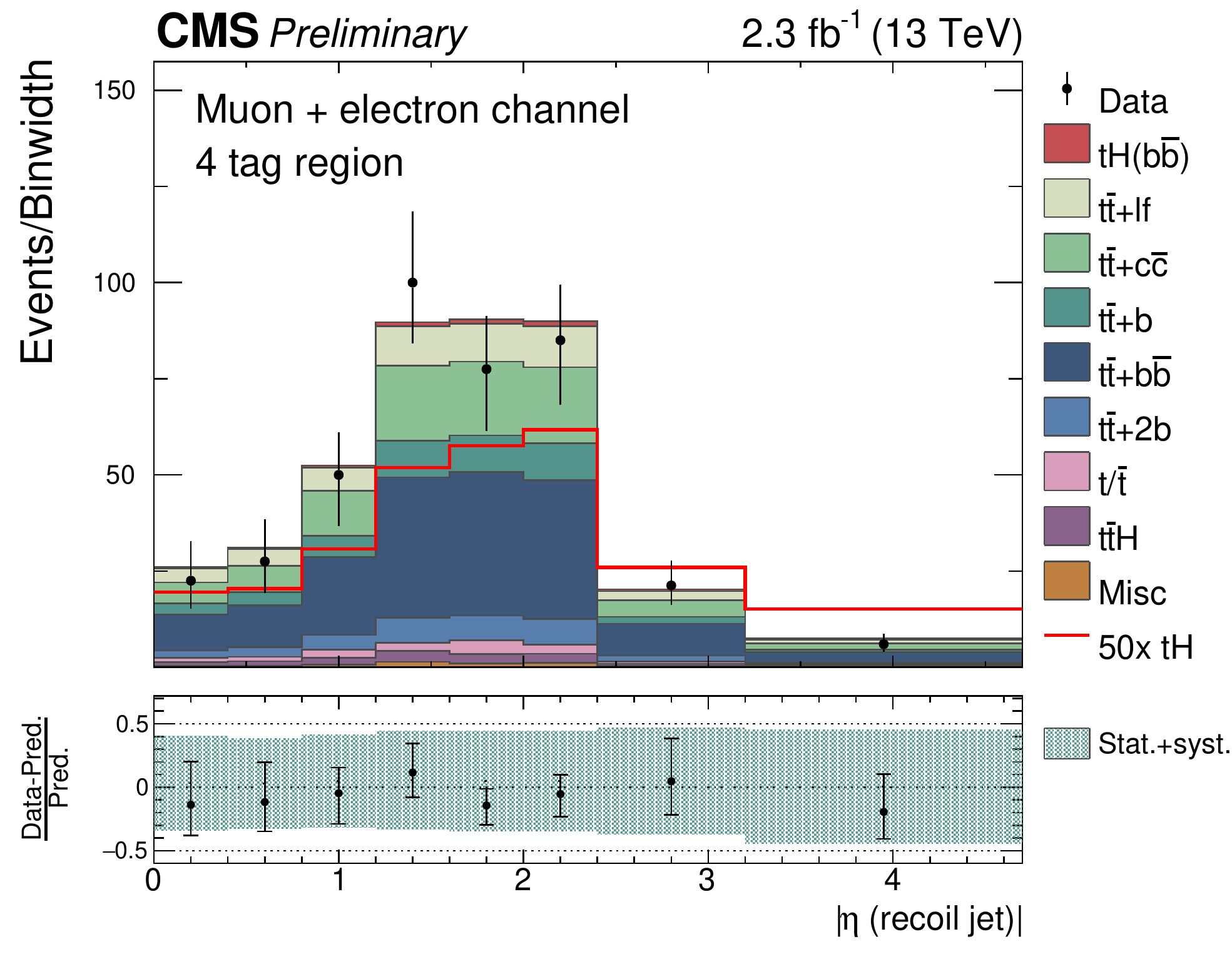}
\caption{Distributions of the two most discriminating classification variables, $\log m(\mathrm{t}_\mathrm{had})$ in the 3 tag region (left) and $|\eta(\mathrm{recoil\ jet})|$ in the 4 tag region (right). The signal distribution is scaled by the factor given in the legends. Taken from~\cite{pas}.}
\label{fig:class_var}
\end{figure}

For the event classification, another 51 BDTs are used. For all BDTs, the signal events (tH production) are trained against the dominating t$\mathrm{\bar{t}}$ and t$\mathrm{\bar{t}}$H background in the 3 tag region. Three types of input variables are used in the training: Variables from one of the 51 tH reconstructions, variables from the t$\mathrm{\bar{t}}$ reconstruction and global variables, which are independent of any reconstruction. The outputs of the classification BDTs are applied to the 3 tag and to the 4 tag regions. In Fig.~\ref{fig:class_var}, the distributions of the two most discriminating variables are shown.

\clearpage

\section{Results}

\begin{figure}[h]
\centering
\includegraphics[width=0.45\textwidth]{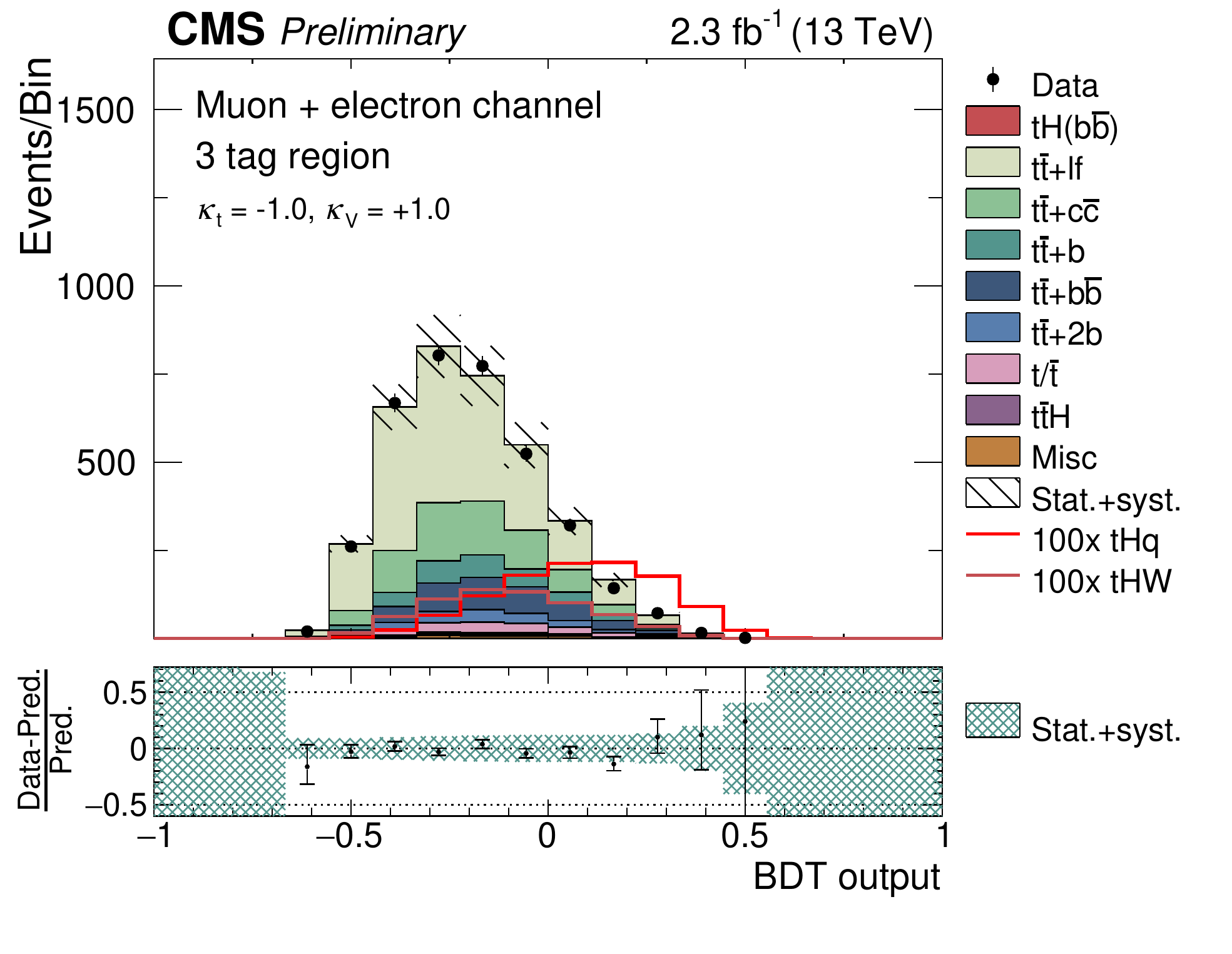}
\hspace*{1.2em}
\includegraphics[width=0.45\textwidth]{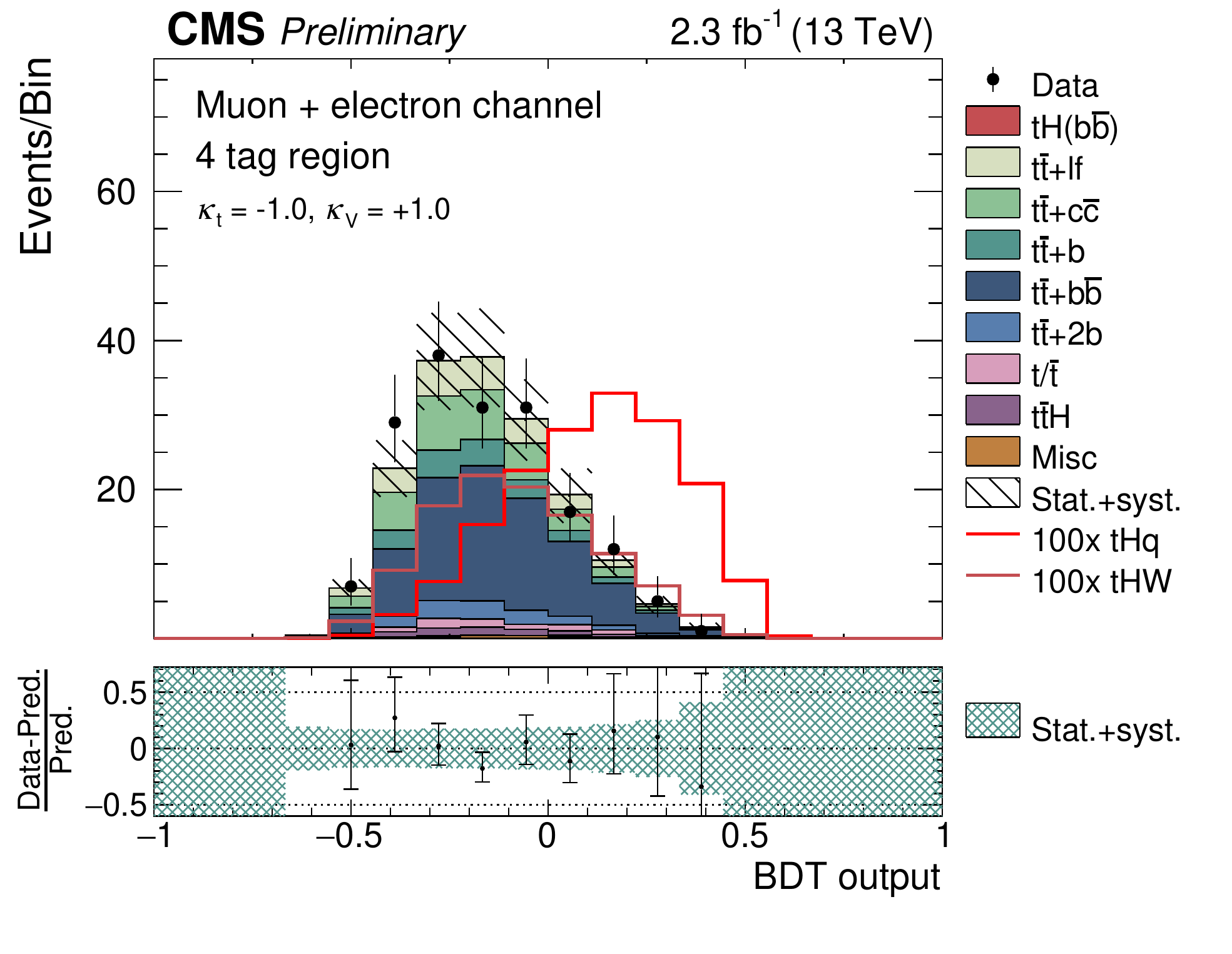}
\caption{Postfit distributions of the classification BDT output for the ITC scenario in the 3 tag region (left) and 4 tag region (right). The signal distributions correspond to the expected contributions scaled by the factors given in the legends. Taken from~\cite{pas}.}
\label{fig:postfit-BDT}
\end{figure}

\begin{figure}[h!]
\centering
\includegraphics[width=0.45\textwidth]{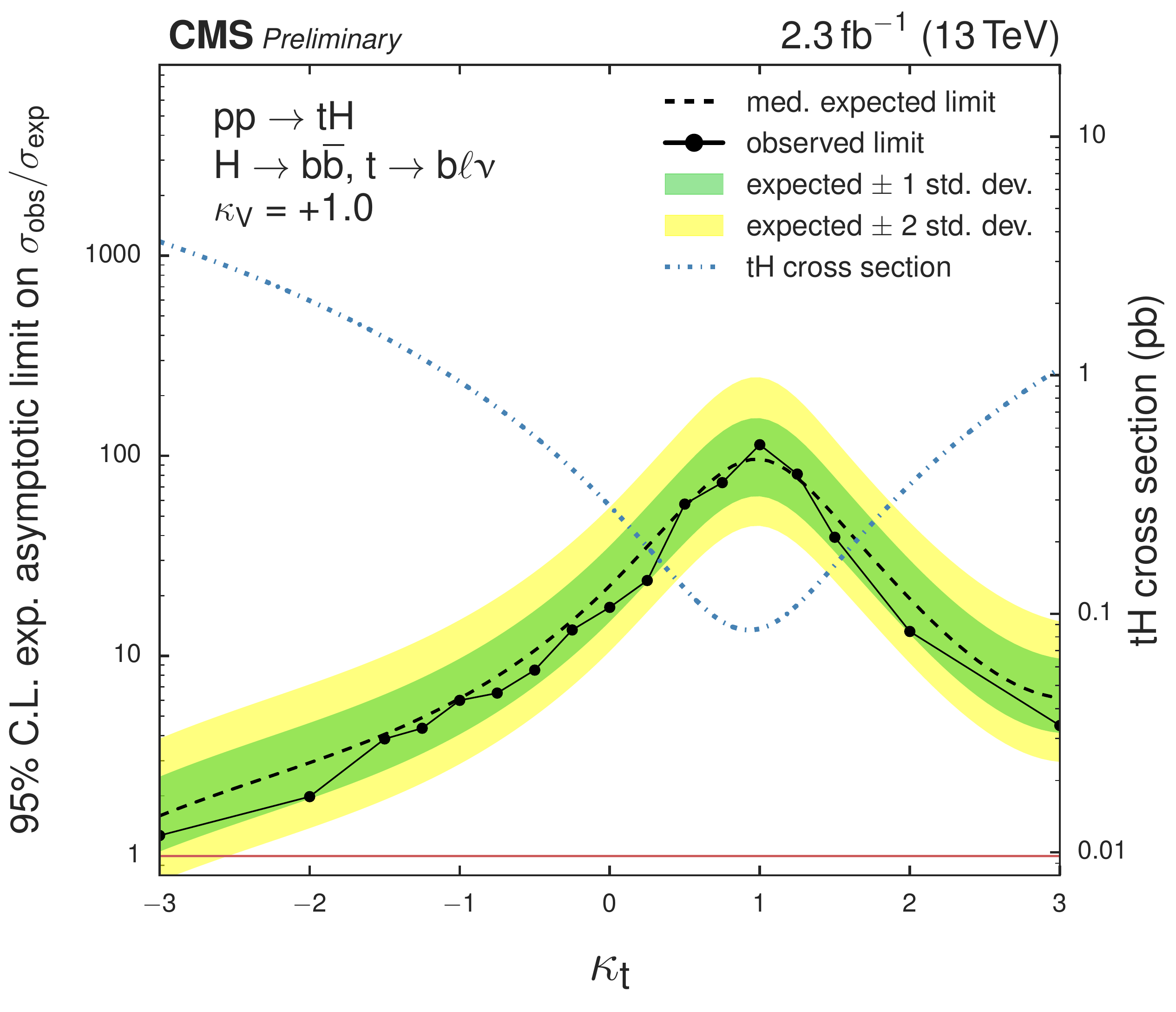}\\
\includegraphics[width=0.4\textwidth]{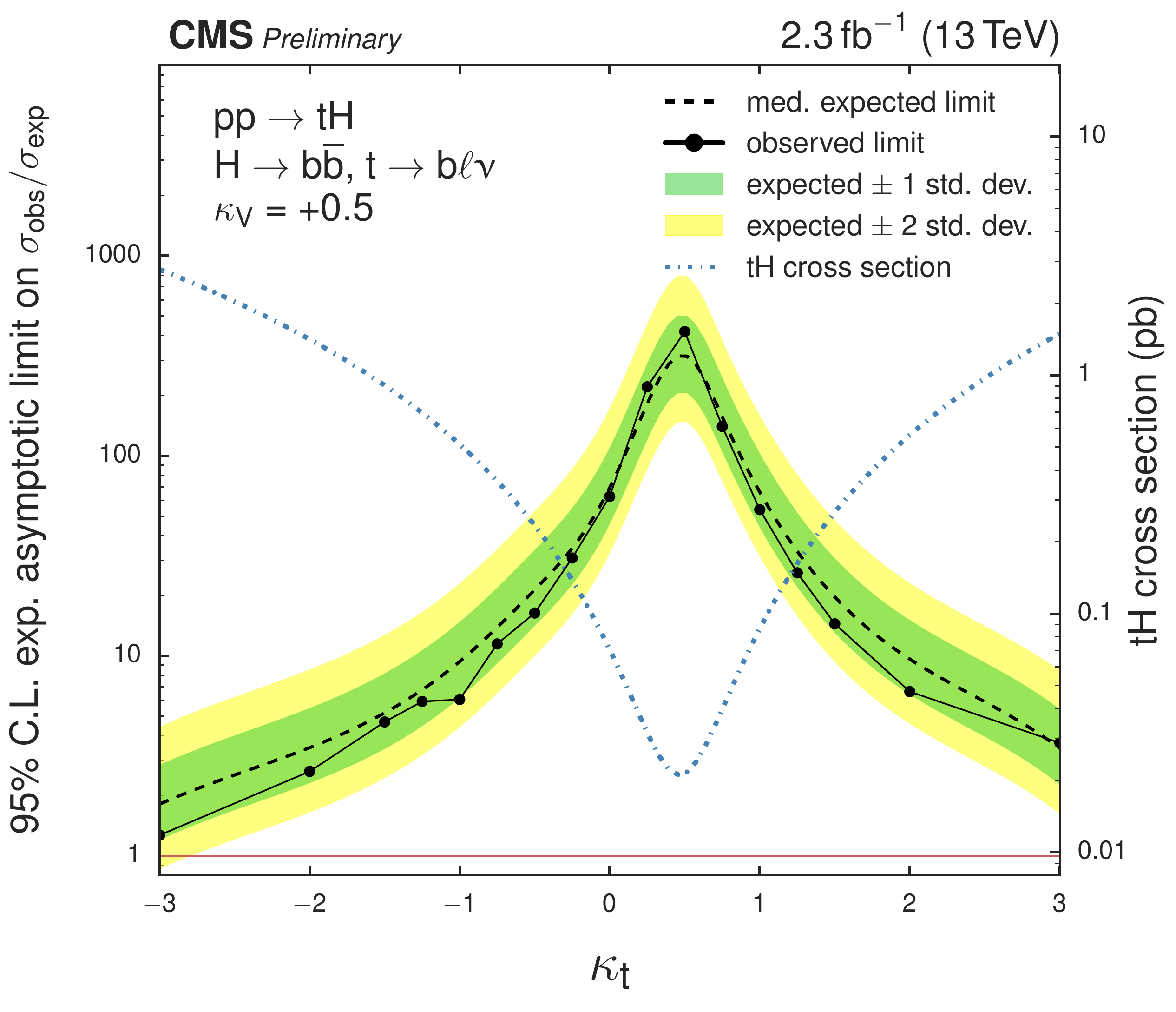}
\includegraphics[width=0.4\textwidth]{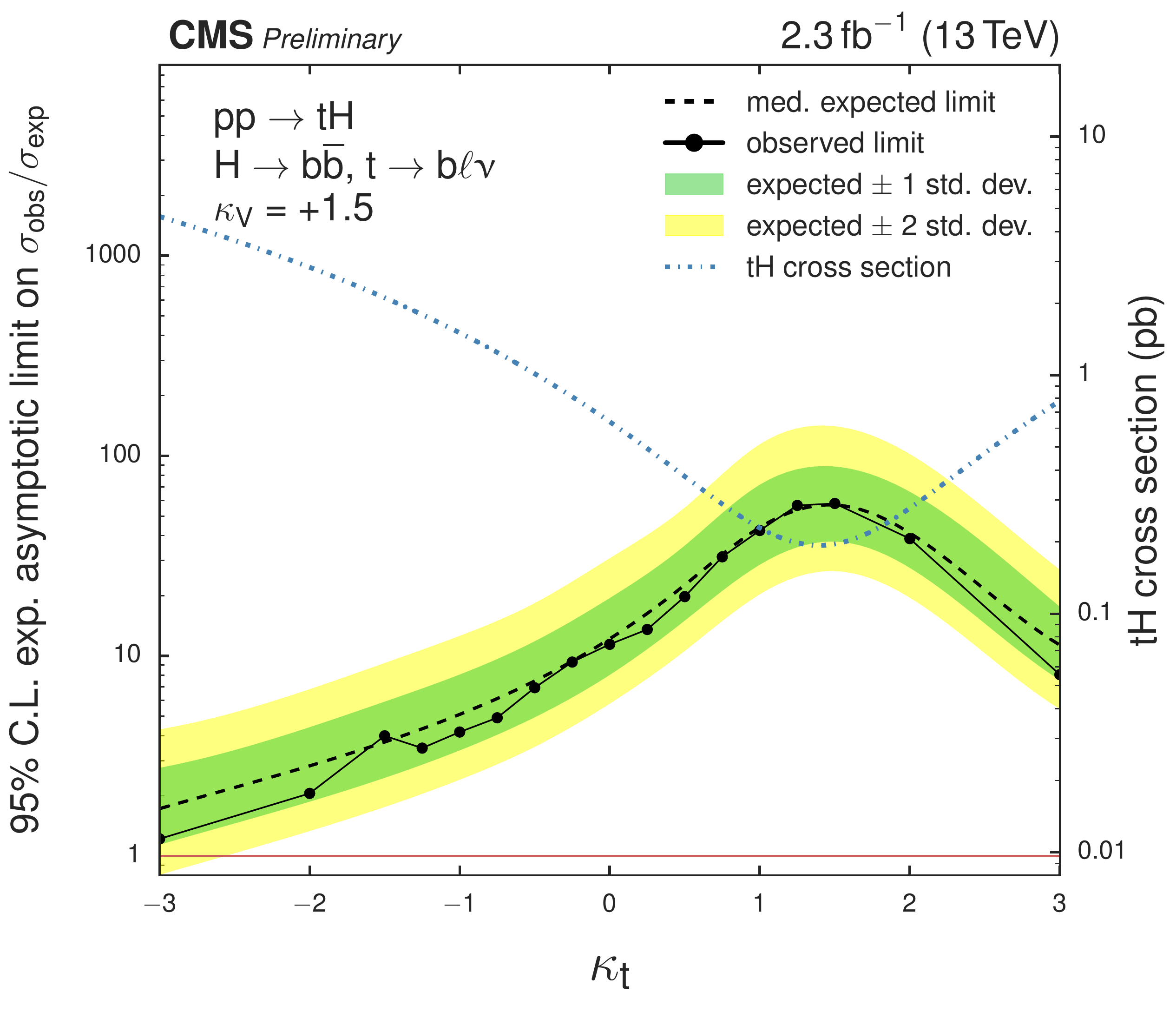}
\caption{Upper limits on tH scenarios with different $\kappa_\mathrm{t}-\kappa_\mathrm{V}$ configurations. Top figure: $\kappa_\mathrm{V}=+1.0$, bottom left figure: $\kappa_\mathrm{V}=+0.5$, bottom right figure: $\kappa_\mathrm{V}=+1.5$. The tH cross sections are given on the right $y$ axis. Taken from~\cite{pas}.}
\label{fig:limits}
\end{figure}

After the event classification, limits for 51 different points in the $\kappa_\mathrm{t}-\kappa_\mathrm{V}$ plane are determined from a simultaneous fit of the corresponding BDT output in the 3 tag and 4 tag regions. The resulting postfit distributions of the classification BDT output for the ITC scenario in the two signal regions are shown in Fig.~\ref{fig:postfit-BDT}. The expected and observed upper limits on the tH production rate for all 51 studied couplings can be found in Fig.~\ref{fig:limits}. For the SM case, the observed limit is $113.7\times\sigma_\mathrm{SM}$ (expected: 98.6), and for the ITC case, an upper limit of $6.0\times\sigma_\mathrm{ITC}$ is observed (expected: 6.4). For all 51 points in the $\kappa_\mathrm{t}-\kappa_\mathrm{V}$ plane, the observed limit is well within one standard deviation of the expected limit. The sensitivity is already comparable to the Run I analysis~\cite{pas_old} which yielded an expected limit of $5.4\times\sigma_\mathrm{ITC}$ for the ITC scenario.

A more detailed description of the analysis is available in Ref.~\cite{pas}.

\clearpage

\end{document}